\documentclass[usenatbib]{mn2e}
\usepackage{epsfig}

\def\equ#1{eq.~(\ref{eq:#1})}

\def\se#1{\S\ref{sec:#1}}

\def\Fig#1{Fig.~\ref{fig:#1}}

\def\beq{\begin{eqnarray}}
\def\eeq{\end{eqnarray}}

\def\kms{\, \rm{km}\,  \rm{s}^{-1}}
\def\rvir{R_{\rm vir}} 
\def\vvir{V_{\rm vir}}
\def\mvir{M_{\rm vir}}
\def\cvir{c_{\rm vir}}
\def\lp{\lambda'}
\def\lpdm{\lambda'_{\rm dm}}
\def\lpb{\lambda'_{\rm b}}

\def\sgl{\sigma_{\lambda}}

\def\omm{\Omega_{\rm m}}
\def\oml{\Omega_{\Lambda}}

\def\drm{{\rm d}}
\def\rb{R_{\rm b}}
\def\rdm{R_{\rm dm}}

\def\jmax{j_{\rm max}}
\def\jtot{j_{\rm tot}}
\def\Jb{J_{\rm b}}
\def\Jdm{J_{\rm dm}}
\def\fbar{f_{\rm b}} 
\def\fd{f_{\rm d}} 
\def\vfb{V_{\rm fb}}
\def\vsat{V_{\rm sat}}

\title[Towards a Resolution of the Galactic Spin Crisis]{Towards a 
Resolution of the Galactic Spin Crisis:\\
Mergers, Feedback, and Spin Segregation}	

\author[Ariyeh H. Maller and Avishai Dekel]
{Ariyeh H. Maller and Avishai Dekel\\ 
Racah Institute for Physics, The Hebrew University,  
Jerusalem 91904, Israel} 
 
\begin{document} 

\maketitle

\begin{abstract} 
We model in simple terms the angular-momentum problems of galaxy formation in 
CDM cosmologies, and identify the key elements of a scenario that may solve 
them.  The buildup of angular momentum is modeled via dynamical friction and 
tidal stripping in a sequence of mergers. We demonstrate how over-cooling in 
incoming halos leads to a transfer of angular momentum from the baryons to the 
dark matter, in conflict with observations. By incorporating a simple recipe of
supernova feedback, we are able to solve the problems of angular momentum in 
disk formation.  Gas removal from the numerous small incoming halos, which 
merge to become the low specific angular momentum ($j$) component of the 
product, eliminates the low-$j$ baryons.  Heating and puffing-up of the gas in 
larger incoming halos, combined with efficient tidal stripping, reduces the 
angular momentum loss of baryons due to dynamical friction. Dependence of the 
feedback effects on the progenitor halo mass implies that the spin of baryons 
is typically higher for lower mass halos.  The observed low baryonic fraction 
in dwarf galaxies is used to calibrate the characteristic velocity associated 
with supernova feedback, yielding $\vfb \sim 100\kms$, within the range of 
theoretical expectations.  We then find that the model naturally produces the 
observed distribution of the spin parameter among dwarf and bright disk 
galaxies, as well as the $j$ profile inside these galaxies.  This suggests that
the model indeed captures the main features of a full scenario for resolving 
the spin crisis.
\end{abstract}


\section{Introduction}
\label{sec:intro}

The `standard' model of cosmology, which assumes hierarchical buildup of 
structure in a universe where the mass is dominated by cold dark matter (CDM), 
seems to be facing intriguing difficulties in explaining some of the robust 
observed properties of galaxies. 
Standing out among these problems is the inability of galaxy formation 
models to reproduce both the sizes and structure of disk galaxies.
In hydrodynamical simulations baryons lose a significant fraction
of their angular momentum leading to overly small disks.
To avoid this, analytic and semi-analytic models commonly assume there is no 
angular momentum loss.  Recent studies, however, show that this leads to 
a variety of discrepancies with observations.
Taken together this suggests a crisis in our understanding of the role 
of angular momentum in forming disk galaxies.
Our aim here is to make progress in the effort to resolve this crisis
by first reproducing the problems using a simple model in which the 
important physical ingredients are spelled out and well understood.  
Subsequently we incorporate an additional process into this model 
which then reproduces the observed sizes and structure of galactic disks.

The sizes of galactic disks are commonly linked to 
the angular momentum of their parent dark-matter halos \citep{fe:80}.  
This modeling is based on the distribution of halo spin parameters 
as found in N-body simulations \citep[see][and references therein]{bull:01b}, 
combined with the assumptions that the baryons and dark matter 
initially share the same distribution of specific angular momentum, $j$, 
within the halos \citep[as seen in simulations by][]{bosch:02}
and that $j$ is conserved as the baryons 
contract to form the disk \citep[as suggested by][]{mest:63}.  
The sizes of disks obtained under these assumptions are roughly comparable
to those observed.

However, high-resolution simulations that incorporate gas
processes find this scenario to be invalid.  In particular, they find that a 
significant fraction of the angular momentum of the baryons is transfered to 
the dark matter, resulting in disk sizes roughly an order of magnitude
smaller then those observed \citep{ns:00,slgv:99,ns:97,nfw:95,nb:91}.
This has been refered to as the {\it angular momentum catastrophe}.
   
The angular momentum catastrophe is commonly 
associated with the problem of ``over-cooling" in CDM-type scenarios
\citep{wf:91,wr:78}.
Without sufficient feedback much of the gas cools quickly,  
contracts into small halos and then spirals deep into the centers 
of bigger halos, efficiently transferring its orbital angular momentum to the 
dark matter \citep{ns:00}.  
It has therefore been speculated that enough energy feedback from 
supernova may prevent this over-cooling \citep{lars:74,wr:78} 
and thus reduce the angular-momentum loss. 
Indeed, simulations where gas cooling is artificially suppressed till $z=1$ 
do not suffer from the angular momentum catastrophe
\citep{eew:00,wee:98}. 
Furthermore, simulations including some forms of feedback show a 
reduction in the amount of angular momentum lost \citep{slgv:99}.
However, while feedback has been studied using simplified approximations 
\citep[e.g.,][hereafter DS]{ds:86},
a realistic implementation of feedback has proved 
challenging \citep[see][and references therein]{tc:00}, though some 
partial progress 
may have been made recently \citep{ft:00,tc:01,sh:02}.
At this point, the feedback scenario has not yet been studied in 
satisfactory detail, nor has it been confirmed to solve the spin problem, 
or properly understood in basic terms.  
This motivates an attempt to understand the scenario and how it may work 
using a simple semi-analytic model, in which the basic elements are 
easily understood. 

Even if the angular momentum catastrophe 
can be avoided such that the total 
baryonic spin agrees with the total dark-halo spin, there 
remain discrepancies between the angular-momentum properties of dark
halos in N-body simulations and those of observed disk galaxies. 
Foremost among these is the apparent mismatch of the 
distribution of specific angular momentum within galaxies
(loosely termed ``the $j$ profile") 
between observations and N-body simulations. 
\citet[hereafter BD]{bull:01b}  revealed a universal profile with an excess 
of both low-$j$ and high-$j$ material compared to the observed
$j$ distribution in exponential galactic disks.  Subsequently, 
\citet*[hereafter BBS]{bbs:01} have measured the $j$ distribution 
in a sample of dwarf disk galaxies, confirming in detail, case by case, 
the differences between the $j$ profile 
predicted by the simulated halos and those found in disk galaxies.
We refer to this second discrepancy as the {\it mismatch of 
angular-momentum profiles}. 

Another spin problem is indicated by the observations of \citet{dl:00}, 
who found that the scatter in the log of the spin parameter 
inferred from fitting the spread of observed disks sizes in a 
large sample of late-type disk galaxies is only $0.36\pm0.03$, while the 
scatter seen in simulated halos is significantly larger, $0.5\pm0.05$ 
(BD and references therein). 
This could have been explained by a scenario in which the disks formed in
low-spin halos are unstable and thus become early-type galaxies, 
which implies that disk galaxies occupy only the higher spin halos 
\citep*{mmw:98,bosch:98}. 
However, this scenario is in apparent conflict with the finding
in N-body simulations that recent major mergers, usually identified with 
large spheroidal components, actually give rise to halos with higher 
spin than average \citep{gard:01,wech:01b}.  This demonstrates that
the exact connection between the angular-momentum properties observed 
in galaxies and in N-body simulations is unclear.

We propose that the solution to the crisis is spin segregation ---
that the angular momentum distribution of baryons differs 
from that of the dark matter due to gas processes. 
These processes
can either decrease or increase the specific angular momentum of the 
baryons relative to the dark matter.  Gas cooling generally results in 
lower spin for the baryons compared to the dark matter, but 
heating due to feedback reduces this effect, and gas removal in small halos 
actually results in a higher spin for the baryons compared to the dark matter.

In this paper we work out a simple model to explore these effects.
Knowing that in a hierarchical scenario the halo buildup can be largely
interpreted as a sequence of mergers between smaller halos, our model is 
based on a simple algorithm for the buildup of halo spin by summation of 
the orbital angular momenta of merging satellites \citep{mds:02,vitv:01}.
This algorithm has been found to match well the spin distribution among
halos in N-body simulations.  In this paper we extend the model
and find that it also reproduces the angular-momentum profile 
within halos. It therefore provides a useful insight into the origin of the
spin problems, and a clue for their possible solution.
We therefore use this model as a tool for incorporating the 
relevant baryonic processes, and especially feedback.

In \se{halos} we describe the model for angular-momentum buildup in halos 
by mergers.  In \se{oc} we model the angular momentum catastrophe 
as resulting from over-cooling and dynamical friction.  
In \se{fb} we introduce a simple model of feedback. 
In \se{obs} we study the resultant angular-momentum properties
of the baryonic component both in bright and dwarf 
disk galaxies and compare them to observations.
In \se{models} we test the robustness of our results to the details
of the feedback model assumed. We conclude and discuss our results in 
\se{conc}.

\section{Buildup of Halo Spin by Mergers}
\label{sec:halos}

\subsection{Modeling the Spin Parameter}

The angular momentum of a galaxy, $J$, is commonly expressed in terms of
the dimensionless spin parameter $\lambda=J\sqrt{|E|}/GM^{5/2}$,  
where $E$ is the internal energy \citep{peeb:69}.  In practice, the 
computation and measurement of this quantity, especially the energy, may 
be ambiguous, e.g., it is not obvious how to define separate spin 
parameters for the dark matter and the baryons.
Furthermore, it introduces an undesired dependence
on the specifics of the halo density profile. 
Instead, following BD, we use the modified spin parameter
\beq
\lp={{j}\over{\sqrt{2}\vvir\rvir}} ,  
\eeq
where $j=J/M$ is the specific angular momentum.
This quantity is more straightforward to compute for each component,
$\lpdm$ and $\lpb$, and 
it does not explicitly depend on the density profile of the halo.  
For a density profile of the type suggested by \citet[NFW]{nfw:96} 
$\lp$ is related to $\lambda$ by 
\beq 
\lp=\lambda f_c^{-1/2} \, ,
\label{eq:lamtolp}
\eeq
where the factor $f_c$ reflects the difference between the energy of an
NFW halo and a truncated singular isothermal sphere.
An expression for $f_c$ as a function of the halo concentration, 
$\cvir$, is given in \citet[][equation 22]{mmw:98}.
For a NFW halo with $\cvir=4.5$ the values of the two spin parameters 
become equal.

BD found for simulated halos that, similarly to $\lambda$,
the distribution of $\lp$ is well fit by a log-normal function,  
\beq
P(\lp)d\lp={{1}\over{\sqrt{2\pi\sgl^2}}} 
\exp{\left(-{{\ln^2{(\lp/\lp_0)}}\over{2\sgl^2}}\right)} 
{{d\lp}\over{\lp}},
\eeq
with $\lp_0 \simeq 0.035$ (compared to $\lambda_0 \simeq 0.042$)
and $\sgl \simeq 0.5$.

In \citet{mds:02}, we proposed a very simple model for the buildup of spin in
dark halos, termed ``the orbital-merger model".
In this model, the spin of a final halo is simply the vector sum of the 
orbital angular momenta of all the halos merging into 
its main progenitor throughout its history.
In this paper, as in \citet{mds:02}, we generate 500 random realizations
of merger histories for each desired final halo mass at $z=0$,
using the method of \citet{sk:99} with the slight adjustment of 
\citet*{bkw:00}
to implement the Extended Press Schechter formalism \citep{lc:93,bond:91}.
The specific cosmology assumed, without significant loss of generality,
is the ``standard" $\Lambda$CDM cosmology, with $\omm=0.3$, $\oml=0.7$, 
$h=0.7$ and $\sigma_8=1.0$. 
For each of the merger trees, we
perform 10 realizations of the orbital angular momentum that
comes in with each merger. The encounter parameters are taken to be
those of typical encounters in simulations, and the directions of the
added spins are drawn at random, allowing for a slight correlation between 
the planes of
successive mergers \citep[as seen in simulations, e.g.][]{dekel:01,pdh:02b}. 
We end up with a distribution of 5000 spin values for each halo mass.
This distribution is shown in \Fig{oc}, in comparison with the log-normal 
distribution obtained by BD from a cosmological simulation. 
The model predictions of $\lp_0=0.036$ and $\sgl=0.53$ match rather well
the simulated results.  One can see in 
Figure 2 of \citet{mds:02} that the match is reasonable even
without introducing any correlation between the encounter planes. 

Similar results were obtained by \citet{vitv:01},
who draw their encounter parameters statistically from 
the distribution found in N-body simulations,
and by \citet*{mtt:02}, where encounter parameters for each object are 
provided by Lagrangian perturbative theory. 
We conclude that the simple orbital-merger model is successful in reproducing
the simulated halo spin distribution, and therefore serves as a 
useful tool for studying the angular-momentum buildup in galaxies.

\begin{figure}  
\centering
\vspace{15pt}
\epsfig{file=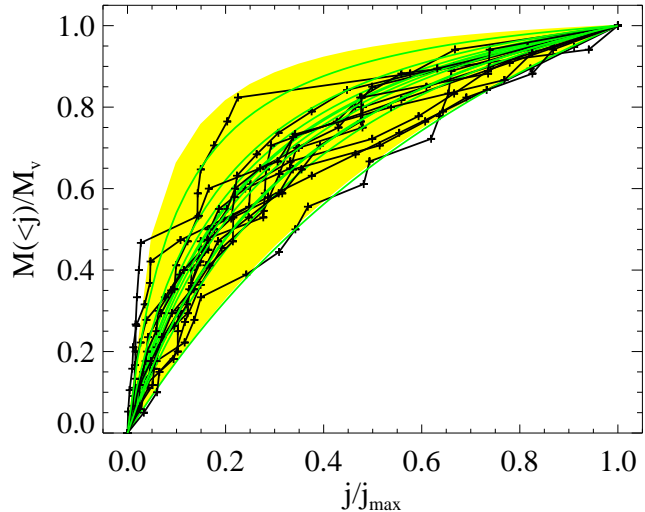,width=\linewidth}
\vspace{10pt}
\caption{Dark-halo angular-momentum profiles: 
The mass fraction with specific angular momentum 
less then $j$ as a function of $j/\jmax$.  
The lines connecting symbols represent a random sample of profiles
produced by our orbital-merger model.  
The smooth curves are fits to these profiles,
where $\mu$ is chosen to cross at half the total mass.
The shaded region marks the $90\%$ spread of profiles found by BD
in simulations.  
}\label{fig:mofj}
\end{figure}

\subsection{Modeling the Angular-Momentum Profile} 
\label{sec:profile}

The cumulative mass distribution of specific angular momentum within 
simulated dark halos, $M(<j)$, was found by BD to be well fit by the 
2-parameter universal functional form
\beq 
\label{eq:mofj}
{{M(<j)}\over{\mvir}}={{\mu\, j/\jmax}\over{\mu-1+j/\jmax}}, \quad \mu>1,
\eeq
in which one of the parameters, say $\jmax$, can be replaced by $\lp$.
This profile is a simple power law, $M(<j) \propto j$, for at least half
the mass. 
When $\mu$ is close to unity, it bends over towards the high-$j$ end.
The distribution of the shape parameter $\mu$ was found to be 
Gaussian in $\log{(\mu-1)}$, with a mean of $-0.6$ 
and standard deviation of $0.4$.

One can also quantify the shape of the angular momentum profile by 
the ratio of the total specific angular momentum, $\jtot$, to $\jmax$.
The relation between this parameter and $\mu$ is 
\beq
\label{eq:zeta}
\zeta={{\jtot}\over{\jmax}}=
1-\mu(1-(1-(\mu-1)\log{\mu\over{\mu-1}}))
\eeq
(BBS eqn. 12).

We create $M(<j)$ profiles for each of the realizations of the
orbital-merger model as follows.
We divide the mass growth of the halo into 20 equal-mass bins 
and assign to each bin the corresponding angular momentum 
that comes in with that mass.
When a satellite's mass is divided among several bins, $i=1,n$,  
with a fraction $f$ in bin 1, we assign a fraction $f^2$ of its 
angular momentum to bin 1. The remaining satellite mass and angular momentum
are then distributed in an analogous way among bins $i=2,n$.
This mimics an $M(<j) \propto j^{1/2}$ profile for the contribution of each
satellite, in rough agreement with the predictions of a toy model based
on dynamical friction and tidal stripping for an NFW halo 
\citep[BD; ][]{dekel:01}.
The role of this procedure is to smooth the profile; in most cases it does
not change the global shape, which is predominantly
determined by the cosmological sequence of mergers.
The result is therefore not too sensitive to the exact assumed distribution 
of $j$ due to each single merger.

\begin{figure}  
\centering
\vspace{15pt}
\epsfig{file=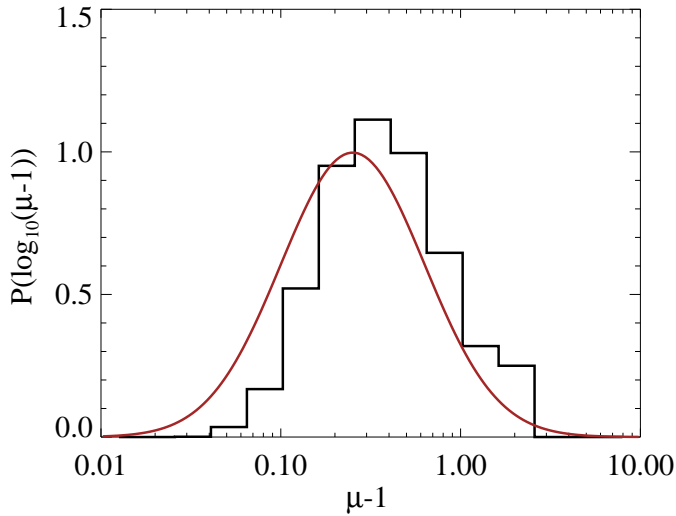,width=\linewidth}
\vspace{10pt}
\caption{The distribution of shape parameter $\mu$ for 
halo angular-momentum profiles. The histogram describes the distribution
in the sample reproduced by our orbital-merger model.  
The smooth curve is the log-normal fit to the N-body simulation 
results of BD.  There is a slight offset in the mean of the distribution
of our model (-0.5 instead of -0.6) but this is not statistically 
significant.
}\label{fig:mu}
\end{figure}

\Fig{mofj} shows a sample of $j$ profiles produced by this procedure.
Our profiles are somewhat noisier then those seen in simulations, 
partly because we ignore any exchange of angular momentum between particles.
The overall agreement between our 
model profiles and the simulation results of BD suggests that
the rearrangement of angular momentum only smoothes the profile 
without changing its global shape.
We fit each of our profiles to \equ{mofj} by the simple method of  
requiring an exact match at half the virial mass (\Fig{mofj}). 
The same results follow if we use $\zeta$ to compute $\mu$ 
by inverting eqn.~\ref{eq:zeta}.
We obtain a distribution of $\mu$ values that are consistent with
the results of BD (\Fig{mu}).  As in BD, they show 
no dependence on halo mass or redshift.
Thus, the orbital-merger model, which reproduces the correct distribution 
of spin values among halos, also recovers reasonably well the angular-momentum 
profile inside halos.

Our orbital-merger model reveals an important
feature in the buildup of halo angular momentum.
We see in the typical realization that
the final halo spin and its direction are predominantly determined by
one last major merger.  Most of the high-$j$ mass in the final distribution
comes in with this merger. The many smaller satellites, each possibly 
carrying a high $j$, come in at different directions and therefore tend to 
sum up to a low spin.  Thus, most of the low-$j$ fraction of the halo mass 
originates in minor mergers.  This feature provides an important clue for 
a possible solution to the spin crisis, as follows.
If small satellites lose a large fraction of their gas before they 
merge into the halo then the final gas distribution in the galaxy
lacks the low $j$ material. Furthermore, if more of the galactic gas 
originates in big satellites then the spin of the baryonic component 
may be higher then that of the dark matter.  
              
\section{Reproducing the Angular Momentum Catastrophe}
\label{sec:oc}

\subsection{Tidal Stripping and Dynamical Friction}

We can understand the transfer of angular momentum from the baryons to the 
dark matter in cosmological hydrodynamical simulations with
a simple model including gas cooling, dynamical friction 
and tidal stripping.  In our model for halo spin buildup discussed above 
we have not specified how the orbital angular momentum of the satellite 
is converted to the spin of the halo. 
Clearly the transfer of angular momentum occurs by the processes of
dynamical friction and tidal stripping.
The dynamical friction acting by the halo particles on the bound part
of the satellite exerts a torque which transfers angular-momentum from the
satellite to the halo and eventually brings the satellite towards the halo
center. Satellite particles that are tidally stripped and become part of 
the halo before the satellite sinks to the halo's center retain 
what is left of 
their angular momentum at that point and add it directly to the halo. 

A hint for the typical $M(<j)$ distribution due to a single merger can be
obtained as follows \citep[BD; ][]{dekel:01}. 
If the satellite moves with a circular velocity $V_{\rm c}(r)$,
and we assume that the lost mass and $j$ are deposited, on average,
locally at $r$, then the resultant spatial profile $j(r)$ can be obtained 
by averaging over shells,
\beq
4\pi r^2 \rho (r)\, j(r)
= m(r){\drm [rV_{\rm c}(r)] \over \drm r}
+{\drm m(r) \over \drm r} rV_{\rm c}(r) \ .
\label{eq:Jdeposit}
\eeq
The first term refers to the $J$ transfer as a reaction to dynamical friction,
and the second is the $J$ deposit associated with the tidal mass transfer.
We need to estimate the momentary surviving satellite mass $m(r)$ in order to
compute $j(r)$, which can then be inverted to obtain the desired $M(<j)$. 

The tidal mass loss at halo radius $r$ can be crudely estimated by evaluating 
the tidal radius $\ell_{\rm t}$ of the satellite at $r$ using the 
crude resonance condition,
\beq
{m(\ell_{\rm t}) / \ell_{\rm t}^3}={M(r) / r^3} \ ,
\label{eq:tidal}
\eeq
where $m(\ell)$ and $M(r)$ are the mass profiles of the satellite and halo
respectively \citep[e. g.][]{wein:94}. 
If, for simplicity, these two profiles are {\it self-similar}, 
then the resonance condition implies
\beq
\ell_{\rm t}/\ell_{\rm vir} =r/\rvir, 
\label{eq:tidal_ss}
\eeq
where $\ell_{\rm vir}$ and $\rvir$ are the virial radii of the satellite and 
the halo respectively.  This means that the remaining bound mass of the 
satellite when it is at $r$ obeys 
\beq
m[\ell_{\rm t}(r)] \propto M(r) .
\label{eq:massloss}
\eeq
A more accurate recipe for tidal stripping, based on studies with 
N-body simulations of mergers, reveal that \equ{massloss} is valid 
as a crude approximation for a wide range of merger parameters 
(Dekel et al., in preparation).  We therefore adopt it in our model.

If the halo density profile is isothermal, $M(r) \propto r$,
then substituting \equ{massloss} in \equ{Jdeposit} yields $j(r) \propto r$,
which corresponds to $M(<j) \propto j$. A more realistic NFW density
profile leads in the outer halo to a $j$ profile closer to 
$M(<j) \propto j^{1/2}$, \citep[BD; ][]{dekel:01} 
which is what we use to smooth the profile in 
the orbital-merger algorithm.  The qualitative similarity between the profile
predicted by this toy model and the typical $j$ profile seen in simulations 
(BD) provides further support to the general picture of $J$ buildup by mergers.

\subsection{Spin Segregation}

We can now explore the effect of cooling in the satellites on the final
angular momentum of the baryons.
We assume that at an early time the baryons follow the dark-matter
distribution. As the gas within a halo cools radiatively, it contracts to
a more compact configuration characterized by
a smaller radius, $\rb$, compared to the extent of the dark matter, $\rdm$.
Such a spatial segregation in a satellite before it merges with a bigger halo
naturally leads to a difference in the spins of baryons and dark matter
in the product halo.  The bound part of the satellite is continuously 
transferring orbital angular momentum to the dark matter by 
dynamical friction. In the extreme case, the inner satellite that remains 
bound all the way to the halo center eventually loses all its orbital angular 
momentum by this mechanism.  At the moment of stripping, the escaping 
satellite material stops being affected by dynamical friction, and it joins 
the halo with the angular momentum that it is left with at that point. 

\begin{figure} 
\centering
\vspace{10pt}
\hspace{25pt}
\epsfig{file=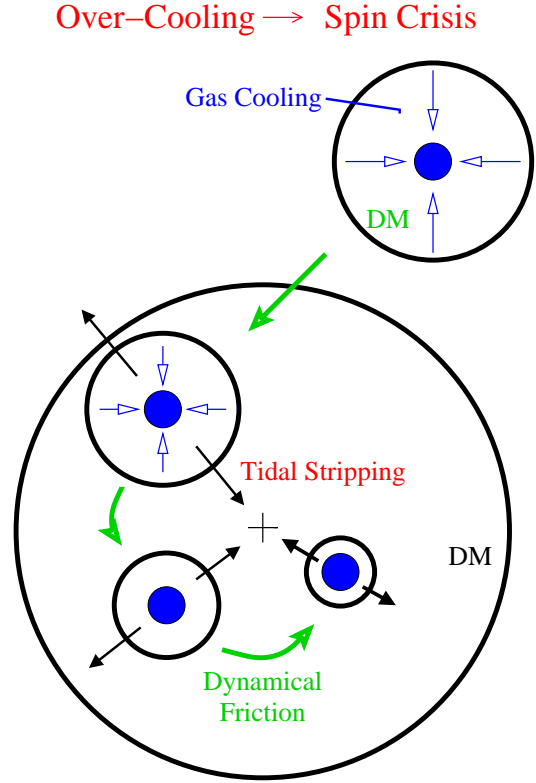,width=\linewidth}
\caption{A schematic illustration of how over-cooling 
in merging satellites leads to the angular momentum catastrophe.  
The gas contraction within the incoming satellite
makes the gas immune against tidal stripping: it spirals all the
way into the halo center while losing all of its orbital angular momentum
to the dark halo due to dynamical friction. The dark matter, which dominates
the outer regions of the satellite, is gradually stripped in the outer
parts of the halo, thus retaining part of its orbital angular momentum.
}
\label{fig:overcool}
\end{figure}

The spatial segregation in the satellite thus implies that the $j$-rich mass
stripped at the early stages of the merger in the outer halo is dominated
by dark matter, while the more compact
baryonic component survives longer as a bound satellite 
and loses more of its orbital angular momentum by dynamical friction.
The result is a net 
transfer from the baryons to the dark-matter component.
This process is illustrated in a schematic diagram, \Fig{overcool}. 

\begin{figure} 
\centering
\vspace{15pt}
\epsfig{file=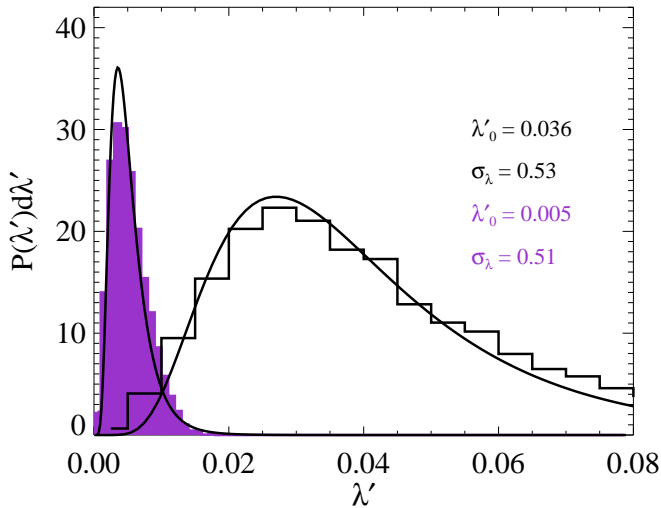,width=\linewidth}
\vspace{10pt}
\caption{The effect of over-cooling on the spin distribution of baryons 
compared to the dark matter. Since the baryons are tightly bound in the 
satellite centers, they spiral into the 
inner part of the big halo without being stripped and thus loose most 
of their angular momentum to the halo by dynamical friction. 
The model reproduces the deficiency of angular momentum in baryons of  
almost an order of magnitude as seen in hydrodynamical simulations.
}\label{fig:oc}
\end{figure}

Using eqn. \ref{eq:tidal_ss} above, we obtain that the 
orbital angular momentum added to the merger product with the baryons 
is related to the amount gained by the dark matter via
\beq
\Delta \Jb ={\rb\over{\rdm}} \Delta \Jdm.
\label{eq:jb}
\eeq
We ignore any transfer of angular momentum from the satellite to the baryons
already in the halo as baryons in the more massive 
halo are already more centrally concentrated than those in the satellite.
In the case of maximum cooling the baryons dominate the center of the halo
which we approximate as $\rb = \fbar\rdm$, 
where $\fbar$ is the universal baryon fraction. 
Adopting $\fbar = 0.13$ \citep{tytl:99}, we obtain that
the spin of the baryons is reduced by almost an order of 
magnitude, and thus reproduces the angular momentum catastrophe
as seen in hydrodynamical cosmological simulations.
In \Fig{oc} we show the resultant probability distribution of the spin 
parameter as obtained in our orbital-merger model with maximum cooling;
it has $\lp_0=0.005$.

This simple modeling of the spin problem makes it easy to understand
why it is usually assumed that some form of energy input into the gas 
may remedy the problem; it would delay the cooling, increase $\rb$,
and thus reduce the baryonic spin loss.

Note that in common semi-analytic models of galaxy formation the gas
is assumed to be either in a hot phase or in a cold phase, 
with the hot phase extending out to $\rdm$ and the cold phase condensed.  
This implies that the cold gas in the satellite 
looses most of its angular momentum to 
dynamical friction, while the hot gas retains most of its angular momentum.
In this case the fraction of baryons in the hot phase, $f_{hot}$, 
determines the orbital angular momentum added as baryonic angular 
momentum to the merger product, 
$\Delta J_{\rm b} = f_{hot} \Delta J_{\rm dm}$.
Thus, one can relate our discussion here to other 
semi-analytic modeling by identifying $f_{hot}=\rb/\rdm$.

\section{Feedback}
\label{sec:fb}

It is generally assumed that some form of heating may play an important role
in preventing over-cooling and therefore angular-momentum loss. 
The most important source of heating is likely to be supernova feedback 
though reionization by the UV background, tidal
heating, and ram pressure may all contribute. Semi-analytic models assume
different ad hoc recipes for the amount of heating as a function of 
star-formation rate, though hydrodynamical simulations have been largely 
unsuccessful so far in implanting feedback in a way that prevents 
over-cooling \citep[see][and references therein]{tc:00}. 
Our approach in the present work is to avoid the details of star formation 
and feedback recipes and rather use a very simple prescription for the 
effect of feedback in a satellite halo as a function of its virial velocity, 
$\vsat$.

Following the analysis in \S IV of DS, we assume that the total supernova 
energy pumped into the gas per unit mass is a robust quantity.
This is based on the assumptions that the supernova rate is roughly 
proportional to the same gas mass that is later being affected by it, 
that the star-formation rate is inversely proportional to the dynamical time, 
and that the supernova remnant transfers most of its energy to the interstellar 
medium by the end of its ``adiabatic" phase.
DS estimated as a crude upper limit that the supernova input is comparable to 
the kinetic energy per unit mass of an isothermal halo with virial velocity
$\sim 100 \kms$.  The actual free parameter characterizing our feedback model,
$\vfb$, is defined as the virial velocity of a halo for which the energy 
input is sufficient to heat all the gas to the virial temperature of the halo. 
Based on DS we expect $\vfb$ to be of order $100 \kms$ or less, but we keep 
this estimate in mind only as a general constraint which does not enter our 
current analysis.  The actual relation between $\vfb$ and the supernova input 
depends on the initial state of the gas before the supernova burst.

We thus parameterize the effect of feedback on the spatial extent of the 
baryons in a halo by the ratio of $\vsat$ to $\vfb$.  
The limit $\vsat >> \vfb$, massive halos with deep potential wells,
corresponds to maximum cooling, $\rb \ll \rdm$.  For smaller halos with 
$\vsat \simeq \vfb$
we expect the heating to balance 
the cooling and yield $\rb \simeq \rdm$. Our model is therefore a simple 
interpolation between these limits,
\beq
\rb=\left({{\vfb}\over{\vsat}}\right)^{\gamma_1}\rdm \hspace{1cm} 
(\vsat > \vfb) \, ,
\label{eq:heat}
\eeq
with $\gamma_1$ an arbitrary exponent.

\begin{figure} 
\centering
\vspace{15pt}
\epsfig{file=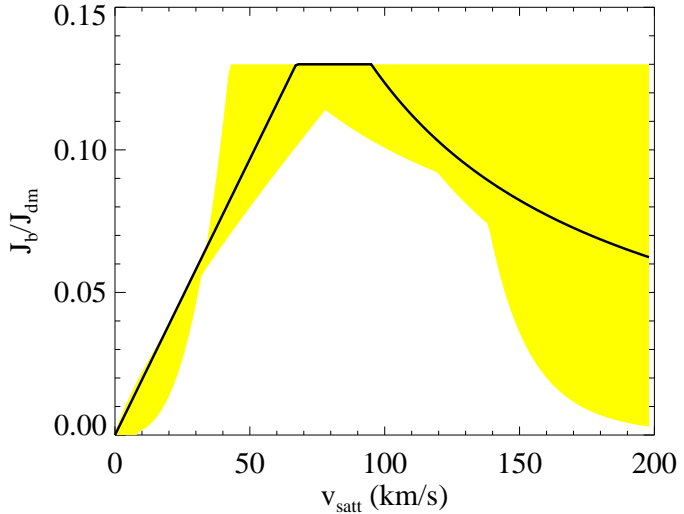,width=\linewidth}
\vspace{10pt}
\caption{
The effect of feedback on the ratio of baryonic angular momentum to that 
of the dark matter as a function of the virial velocity of the merging
satellite. In the model described by the thick curve,
the feedback is characterized by $\vfb=95 \kms$,
with the exponents $\gamma_1=\gamma_2=1$.
If the baryons in the satellite are distributed like
the dark matter (middle range), then $\Jb/\Jdm=\fbar$, the 
cosmological baryon fraction.
In larger satellites with $\vsat > \vfb$,
the baryons are less extended then the dark matter and thus loose more
angular momentum by dynamical friction before they are stripped into the
halo. In smaller satellites with $\vsat < \vfb/\sqrt{2}$,
gas is blown out and therefore $\Jb$ is reduced, though the specific 
angular momentum of the baryons remains the same as that of the dark matter.
The shaded region shows the range of models we consider in 
\S~\ref{sec:models}, all constrained to match the baryonic fraction
in dwarf galaxies.}
\label{fig:models}
\end{figure}

If $\vfb$ is larger than $\vsat$, the feedback can cause gas ejection from
the satellite.  For $\vsat << \vfb$
we expect total blowout,
and we assume that partial blowout starts occurring for halos
where the potential energy of the gas is comparable to the energy input 
from supernova, $\vsat^2 = (1/2) \vfb^2$. 
We therefore parameterize the amount of gas that remains in the halo by
another simple interpolation,
\beq
\fd=\left({{\vsat}\over{\sqrt{2}\vfb}}\right)^{\gamma_2} \hspace{1cm} 
(\vsat < \vfb/\sqrt{2}) \, ,
\label{eq:blowout}
\eeq
with $\gamma_2$ another arbitrary exponent.

\begin{figure*} 
\centering 
\vspace{-75pt}
\epsfig{file=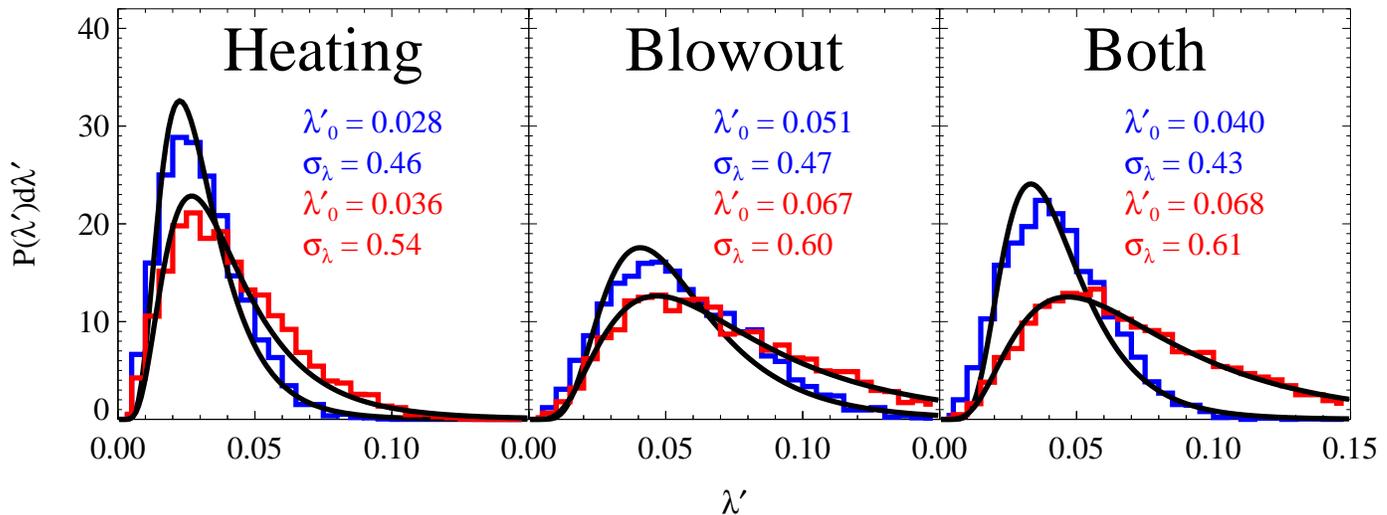,width=\linewidth}
\vspace{-65pt} 
\caption{The effects of heating and blowout in our
model on the distribution of baryonic spin parameter $\lp$.
Log-normal fits are shown for each histogram, with the
corresponding mean and scatter quoted.  In each panel the higher peaked curve
and higher two numbers are the bright galaxies ($\vvir=220 \kms$) and
the lower peaked curve and lower two numbers are the dwarf galaxies
($\vvir=60 \kms$).
{\bf Left}: 
Only heating is included in our feedback recipe (blowout ignored).
{\bf Center}: 
Only blowout is included (cooling ignored).
{\bf Right}:
The whole feedback recipe is included, with cooling, heating and blowout.
}
\label{fig:3lam}
\end{figure*}

We start with the simplest arbitrary choice $\gamma_1=\gamma_2=1$. Later,
in section \ref{sec:models}, we will explore the robustness of the results 
to different assumptions about choices of values for $\gamma_1$ and $\gamma_2$.

The only remaining free parameter of the feedback model is $\vfb$. 
We set it by requiring that halos with a virial velocity of $60 \kms$
have an average $\fd=0.04$, as measured in the data of BBS.  
Note that we are concerned with the state of the baryons 
only when the satellite mergers with the halo;
it is therefore sufficient that the feedback be effective for a short
duration, perhaps triggered by the merging process itself. 

The effects of heating and blowout on the angular momentum 
of baryons versus dark matter, according to our model,
in shown in Figure \ref{fig:models}. 
For massive satellite with 
$\vsat > \vfb$, 
the baryons retain less angular momentum
as the mass increases, because of dynamical friction.
Small satellites with 
$\vsat < \sqrt{2}\vfb$ 
allow a blowout of a larger fraction
of their gas as the mass decreases, thus adding to the merger product less
baryons and therefore less baryonic angular momentum, while the specific
angular momentum of the two components remains the same.

In \Fig{3lam} we demonstrate the effects of the different ingredients
of this feedback scheme, with $\vfb=95 \kms$, on the baryonic spin-parameter
distribution. We do it for two kinds of final halo 
masses, corresponding to virial velocities of $60$ and $220\kms$.
We refer to these as representing {\it dwarf\,} and
{\it bright\,} galaxies respectively.

The left panel of \Fig{3lam}
shows the effects of adding heating only, ignoring 
blowout.  The dwarf galaxies are all built up by satellites of 
$\vsat < \vfb$, 
so there is full heating, $\rb \simeq \rdm$ 
The result is that the spin distribution of the baryons resembles that of 
the dark-matter.
In bright galaxies, the partial heating raises the baryonic spin 
to $\lp_0=0.028$, which is still lower than the typical
spin of the dark matter due to the partial cooling.

The center panel shows the effects of full heating ($\rb \simeq \rdm$) 
and blowout, as if there is no cooling. 
For the two kinds of galaxies, the baryonic $\lp$ distribution
becomes significantly higher then that of the dark matter,
with $\lp_0=0.051$ for the bright galaxies and a very high $\lp_0=0.067$ 
for the dwarf galaxies. 

Finally, the right panel shows the result of applying the full
feedback scheme, considering cooling, heating and blowout together.
The baryons in the bright galaxies now have spins comparable and 
even slightly higher then their dark-matter halos, with 
$\lp_0=0.040$ (and $\sgl=0.43$), while the 
baryons in dwarf galaxies now have spins higher by $\sim 50\%$,
with $\lp_0=0.068$ (and $\sgl=0.61$). 

Thus, the baryonic spin in dwarf galaxies, which are made by mergers
of small satellites, is dominated by the blowout of the gas from these
satellites, and it ends up with $\lpb > \lpdm$. For bigger galaxies, 
which are largely made of bigger satellites, the dominant effect 
preventing the over-cooling spin catastrophe 
is the heating, with some contribution from blowout. 
This yields a baryonic spin distribution similar to that of the dark matter
halos, in general agreement with observations.

\section{Model versus Observations}
\label{sec:obs}

We now confront the model predictions with observations,
in particular those of BBS.
We first calibrate the value of the one free parameter $\vfb$
by the observed gas deficit in dwarf galaxies.  
We then address the distribution of the baryonic spin parameter in these 
galaxies, as well as the angular momentum profiles within dwarf and bright 
disks.

\subsection{Observations}

Generally, galaxy formation modelers have 
compared their predicted disk sizes to the distribution of disk scale 
lengths found in large samples of late type galaxies \citep{cour:96,mfb:92}.
A more detailed comparison has been performed by \citet{dl:00} where they
constructed the bivariate space density of galaxies as a function 
of luminosity and size.  This allows the spread in observed disk sizes 
to be compared to models and they found that the spread corresponded 
to a $\sgl$ of only $0.36\pm 0.03$. Ideally one would like to compare 
the actual angular momentum in models with observed galaxies instead 
of just properties derived from the angular momentum. 

The only observational attempt so far to measure spin parameters for galactic
disks is by BBS, which therefore serves as the main target for our 
modeling effort in this paper.
BBS used fits to the rotation curves and assumed an NFW profile for the
halos \citep{nfw:95} in order to determine the halo virial quantities
$\rvir$ and $\vvir$ for 14 dwarf galaxies, with an average
$\vvir \simeq 60\kms$.
They also assumed a mass-to-light ratio of unity in the $R$ band, 
but their results are not very sensitive to this assumption
because dwarf galaxies are dominated by their dark matter halos. 
BBS then determined for each galaxy the baryonic spin parameter 
$\lambda_{\rm b}$ which we have converted to $\lpb$ using their values
for $\cvir$ and equation \ref{eq:lamtolp}.  This leads to  
an average value of $\lpb \simeq 0.063$, significantly larger than that 
of simulated dark-matter halos.
They also estimated the ratio $\fd$ of disk mass (stars + gas) to 
dark-matter mass, and found an average of $\fd \simeq 0.04$, about a factor
of 3 smaller than the universal baryonic fraction \citep{tytl:99}.
Finally, they measured the $j$ distribution 
in each galaxy, and confirmed the mismatch of angular-momentum profiles. 
We discard one galaxy (UGC4499) from the sample because its rotation-curve 
fit yields strikingly anomalous results. 

\begin{figure}  
\centering 
\vspace{15pt} 
\epsfig{file=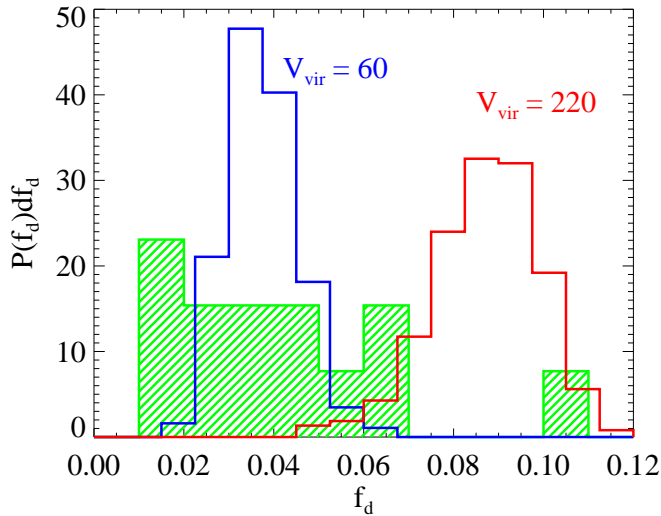,width=\linewidth}
\vspace{10pt} 
\caption{The probability distribution of the fraction of mass in  
baryons $\fd$ for the data of BBS (shaded), in comparison
with our model predictions with $\vfb=95\kms$
for dwarf galaxies (left) and for bright galaxies (right).
The model prediction for the dwarf galaxies, 
with significant blowout, is in good agreement with the BBS data.
The bright galaxies retain most of their baryons.
}\label{fig:fd}
\end{figure}

\subsection{Baryon Fraction}

The distribution of $\fd$ values for the dwarf galaxies of BBS
is displayed in \Fig{fd} as a shaded histogram. These values are 
significantly lower than the generally adopted
universal value of $\fbar \simeq 0.13$, which suggests 
significant baryonic mass loss from these objects. 
Shown for comparison are the corresponding model predictions for
dwarf and bright galaxies. We have chosen a value of $\vfb = 95 \kms$
in order to match the mean of the BBS measurements 
for dwarf galaxies, $\fd \simeq 0.04$, but one can see in \Fig{fd}
that the resulting spread in $\fd$ values is also in good agreement with 
the dwarf galaxies data.
For bright galaxies, $\fd$ is typically
lower than the universal value by only $\sim 30-40\%$, reflecting the
limited fraction of small merging satellites who lost 
their gas.\footnote{
There is one dwarf galaxy with an $\fd$ value much higher then the rest of 
the distribution. This galaxy has the lowest virial velocity
in the sample, $\vvir=43 \kms$, and the second highest concentration 
parameter, $\cvir=31$, 
suggesting that there may be something unusual about this galaxy, either 
in its history or in the BBS fit to it rotation curve.}
It is encouraging that the obtained value for $\vfb$ is in the 
range of expected values for supernova-driven winds (DS).

\begin{figure} 
\centering 
\vspace{15pt}
\epsfig{file=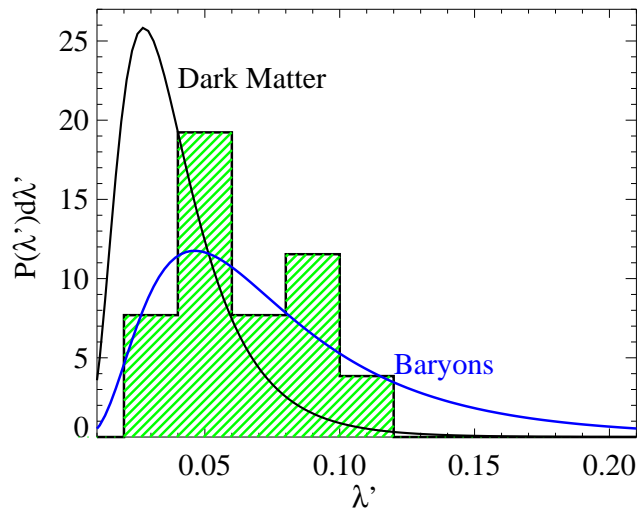,width=\linewidth}
\vspace{10pt} 
\caption{The distribution of $\lp$ in the sample of dwarf galaxies by BBS
(shaded histogram) compared with our model predictions with
$\vfb=95\kms$ for the baryons in dwarf galaxies 
(curve on the right), and the simulation result
for dark halos by BD (curve on the left).
The inclusion of blowout produces a $\lp$ distribution in good 
agreement with the data. 
}\label{fig:data}
\end{figure}

\subsection{Spin Parameter of Baryons}

Next, we compare predicted and observed distributions of spin parameters 
for dwarf galaxies. 
The distribution of $\lpb$ for the BBS disks of dwarf galaxies  
is plotted in Figure \ref{fig:data} as a shaded histogram.
Shown first in comparison is the $\lp$ distribution for dark halos in 
cosmological $\Lambda$CDM simulations from BD.  Contrary to comments 
made by BBS, these two distributions are not in agreement --- 
the observed spins are significantly higher,
with an average of $0.063$ compared to $0.035$ in the dark-matter simulations.
Finally shown in Figure \ref{fig:data} is our model prediction for the 
baryonic spin distribution in dwarf galaxies ($\vvir=60 \kms$), 
with $\vfb=95\kms$.
The effect of blowout brings the baryonic spin distribution into 
good agreement with the observed sample. 
Note that this distribution is quite different then that of the dark matter, 
with $\lp_0=0.068$ and $\sgl=0.61$.
We find it remarkable that the agreement with the observed spin distribution
follows automatically from the adjustment of $\vfb$ to match the observed
baryon fraction in dwarf galaxies.
In the next section we demonstrate that the result is also
independent of the details of the feedback model used. 

\begin{figure*} 
\centering 
\vspace{15pt}
\epsfig{file=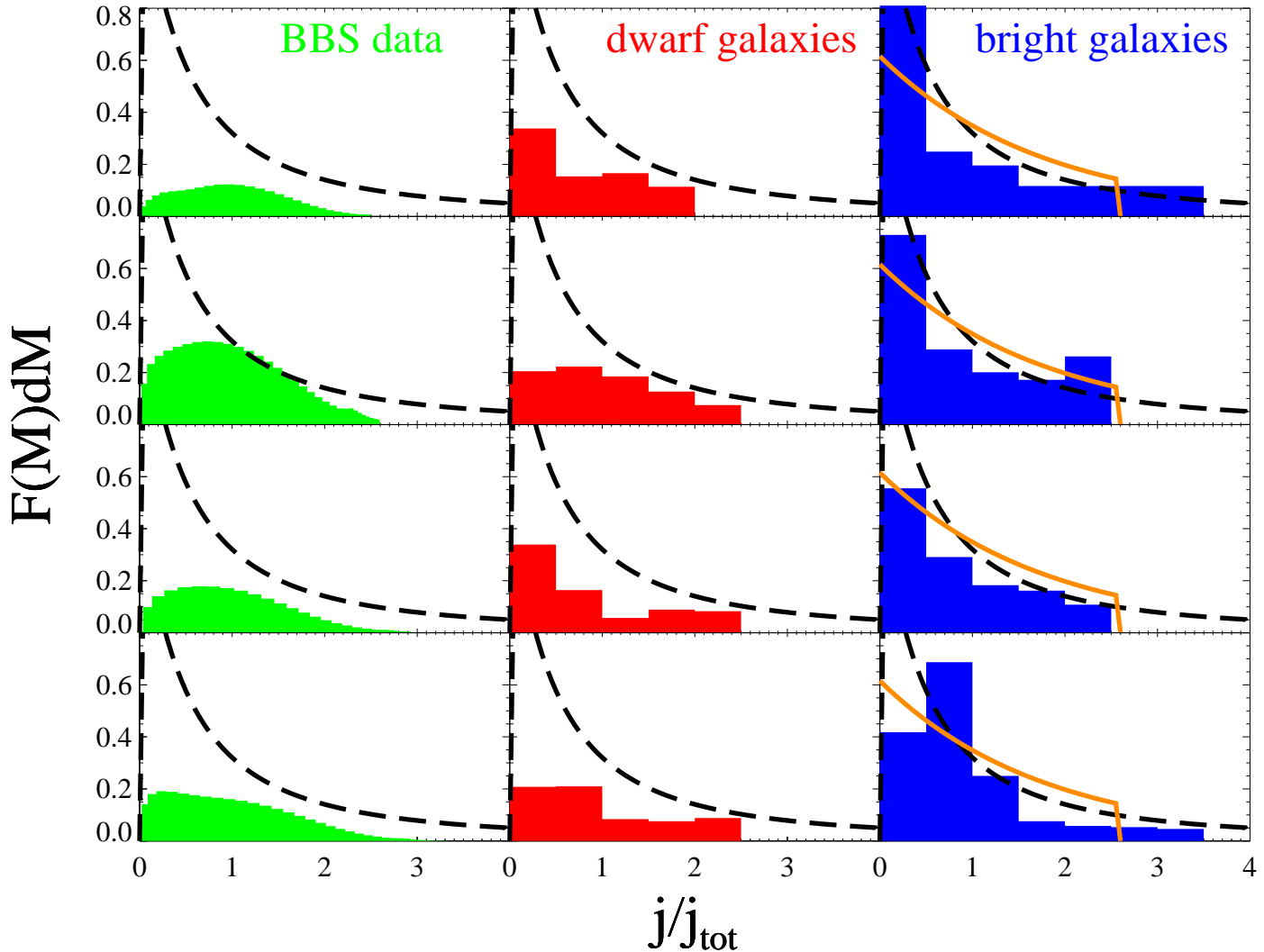,width=\linewidth}
\vspace{20pt} 
\caption{Examples of baryonic specific angular-momentum profiles 
for the BBS data (left column) and for our model dwarf 
and bright galaxies (center and right columns respectively). 
Also drawn is the average ($\mu=1.25$) dark-matter 
$j$-profile seen in the simulations of BD (dashed curve). 
The BBS profiles clearly lack the high and low $j$ tails seen in the 
simulations.  Likewise, many of our model dwarf galaxies also lack 
this material because of gas loss. The bright galaxies, which
retain most of their baryons, have $j$ profiles more similar to the 
dark matter.  However, in some cases they are consistent with the profile of 
an exponential disk with a flat rotation curve (solid curve in right column).
}\label{fig:amd}
\end{figure*}

In the context of the scatter issue mentioned above 
we note that the predicted spread in $\lp$ values for baryons in 
bright galaxies is smaller than the spread for the dark-matter halos 
($\sgl=0.43$ versus $0.5$), and is more like the spread
observed by \citet{dl:00}. We note also that the high baryonic spin 
values obtained in our model for both dwarf and bright 
disks suggests that one does not need to eliminate low-spin halos 
in order to match the observed distribution of disk sizes
[e.g., as done in \citet{mmw:98}].

\subsection{Angular-Momentum Profile of Baryons} 
\label{sec:bprofile}

BBS also measured the specific angular momentum profiles of their dwarf
galaxies, which they find to be quite different from the profiles 
of dark halos found by BD.  
The dwarf galaxies in their sample show a systematic behavior, with
a low baryonic fraction,
and a significant deficit of angular momentum at the two ends of 
the distribution compared to the halo $j$ profiles of BD. 
Examples of some of the $j$ profiles from their data are shown in 
the left column of \Fig{amd}.
 
We construct a baryonic $j$ profile in each of our model realizations,
following the same method used to produce dark-matter $j$ profiles 
in \se{profile}, but now including feedback and gas loss.
Examples of such model profiles for dwarf galaxies of $\vvir=60\kms$
are shown in the central column of \Fig{amd}.  
The low-$j$ tail of the halo distribution is missing in the baryons
in many cases, as expected, because the blowout preferentially 
removes gas from small satellites --- those satellites that merge 
to form the low-$j$ halo material. The high-$j$ tail tends to be reduced
in the baryons, because this tail is often the result of a small satellite 
that comes in with its orbital angular momentum aligned with the halo spin, 
and now has lost its gas.
While there is general agreement between the model dwarf galaxies and the 
BBS data, sometimes the model galaxies show a spike of low $j$ material 
that is not seen in the data.  
We can think of several alternative reasons for this apparent discrepancy.
Either the predicted spike corresponds to a baryonic component that 
BBS fail to observe (such as faint halo stars), or these galaxies
which show a spike do not become disk dwarf galaxies and are therefore
missing from the BBS sample, or our model is missing a certain element that
should eliminate the very low-$j$ spike more efficiently.

\begin{figure}  
\centering 
\vspace{15pt} 
\epsfig{file=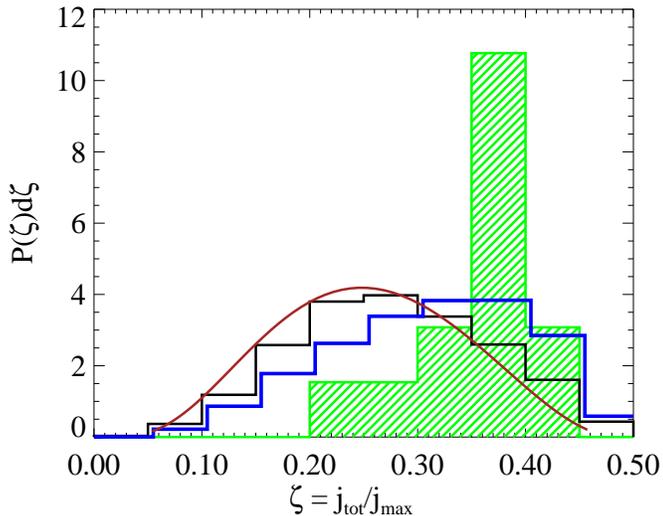,width=\linewidth} 
\vspace{10pt} 
\caption{Probability distribution of the spin-profile shape parameter 
$\zeta=\jtot/\jmax$ (the BBS alternative to $\mu$ of BD) 
for the dwarf-galaxy data of BBS (shaded histogram) 
in comparison with our model predictions for the baryons in dwarf
(thick histogram) and bright (thin histogram) galaxies.  
Shown also is the result for simulated halos by BD (smooth curve).
The model prediction for baryons is in reasonable agreement with the data,
though the predicted spread is somewhat larger.  Bright galaxies show 
a slightly shifted distribution to higher values of $\zeta$.
}\label{fig:zeta}
\end{figure}

The right column of \Fig{amd} shows the model predictions for a sample of 
bright galaxies, $\vvir=220\kms$.  They retain most of their baryons, so 
their profiles are less affected by blowout, 
and they are more similar to the dark-matter $j$-profiles seen in the
simulations of BD.  However, in some cases they show a
reduction of both low and high $j$ material for the same
reasons that it is seen in the dwarf galaxies. 
For comparison, we also show in the right column of \Fig{amd}
the $j$ profile of an exponential disk with a flat rotation
curve truncated at $4.5$ disk scale lengths.  
The $j$-profile mismatch problem in bright galaxies is partly
remedied by the common presence of low-$j$ bulge components in such galaxies.
The fact that sometimes the low-$j$ material is lost even from bright 
galaxies allows some of them to become pure exponential disks, as observed,
without the need to hide any low-$j$ material in a bulge.

\Fig{zeta} shows the distributions of the quantity $\zeta$,
which characterizes the shape of the $j$-profile.
It compares the observed values in the BBS dataset
with our model predictions for the baryons in dwarf and in bright galaxies.
Given the crudeness of our model, the agreement between our model for 
dwarf galaxies and the data is quite reasonable.
The somewhat smaller spread in the data compared to the model predictions 
may suggest that either
selection effects or gas processes tend to favor a particular value of $\zeta$.
Also shown for comparison is the distribution of $\zeta$ for dark halos as
measured by BD in simulations.  One sees that the values of $\zeta$ for 
bright galaxies are slightly shifted to higher values then the dark halos.
For dwarf galaxies the shift is a significant effect, reflecting
the mismatch between the $j$-profiles of dark halos and observed galaxies.

\section{Robustness of the Feedback Model}
\label{sec:models}

In this section we examine the sensitivity of our results in the previous
section to our extremely simple model of feedback. 
We explore a range of values for the exponents $\gamma_1$ and $\gamma_2$.
For each choice of exponents, we determine $\vfb$ such that the mean value of
$\fd$ for the dwarf galaxies is the same as the BBS value of $0.04$ 
We find that the results of \S~\ref{sec:obs} for dwarf galaxies remain
practically unchanged.
The results for bright galaxies do depend on the value of $\gamma_1$, 
though a value can always be chosen such that the median spin
for the bright galaxies is at least as high as that of the dark halos. 

In Table \ref{tab:mod} we show a range of values of $\gamma_2$ and 
the corresponding values of $\vfb$ necessary to give our dwarf halos the
right baryon fraction $\fd$.  Then we list the largest value of $\gamma_1$ for 
which bright galaxies have a median $\lpb \ge 0.035$, as suggested 
observationally. 
A maximum value of $3.0$ is considered for both exponents as 
higher values produce indistinguishable results. 
The resulting values of $\fd$, $\lp_0$ and $\sgl$ are listed
for the two types of galaxies.
We indeed see in the table that varying the exponents 
$\gamma_1$ and $\gamma_2$ has a small effect on the resulting 
baryonic spin distribution for dwarf galaxies. We conclude that any feedback
recipe in which an appropriate amount of gas is being ejected from dwarf 
galaxies should resolve the angular-momentum crisis 
for dwarf galaxies
in a way similar to our results in \se{obs}.  The angular momentum catastrophe 
for bright galaxies is prevented if $\gamma_1$ is sufficiently small.

If $\gamma_2$ is as small as $0.5$, as in the first row of 
Table \ref{tab:mod}, then the required value of $\vfb$ needed to fit 
the mean value of $\fd$ in dwarf galaxies is very high ($185 \kms$).
This velocity is significantly higher than the crude upper limit
allowed by the energetics (DS).  Also, for such a large $\vfb$ value 
bright galaxies have a median $\lp$ much higher then the dark matter 
even for the maximum value of $\gamma_1$.
Thus, the model yields sensible results only if $\gamma_2$ is limited
to the range $0.8-3.0$. 

The shaded region in \Fig{models} shows the range of curves
corresponding to the range spanned in Table \ref{tab:mod}.
We note that a tight constraint is enforced by the data on this range of 
feedback models, that halos of $\sim 35 \kms$ must loose about half of their
baryons.

\begin{table}
\caption{The dependence of our results on the model parameters.  For
a given $\gamma_2$ the scale $\vfb$ is set by requiring the average 
$\fd$ in dwarf galaxies ($\vvir=60 \kms$) to be $0.4$.  Then we find 
the lowest value of $\gamma_1$ that creates a spin distribution in 
bright galaxies ($\vvir=220 \kms$) with a median $\geq 0.035$.}
\label{tab:mod}
\begin{tabular}{ccccccccc}
\multicolumn{3}{c}{\underline{model}} & 
\multicolumn{3}{c}{\underline{dwarf galaxies}} & 
\multicolumn{3}{c}{\underline{bright galaxies}}\\
$\gamma_2$ & $\vfb$ & $\gamma_1$ & 
$\fd$ & $\lp_0$ & $\sgl$ & $\fd$ & $\lp_0$ & $\sgl$\\
\hline
0.5 & 185 &  3.0 & 0.4 & 0.061 & 0.56 & 0.8 & 0.053 & 0.50\\
0.8 & 130 &  3.0 & 0.4 & 0.063 & 0.56 & 0.8 & 0.035 & 0.53\\
1.0 &  95 &  1.5 & 0.4 & 0.067 & 0.61 & 0.9 & 0.035 & 0.44\\
3.0 &  60 &  0.5 & 0.4 & 0.082 & 0.64 & 0.9 & 0.036 & 0.42\\
\hline
\end{tabular}
\end{table}

\section{Conclusion} 
\label{sec:conc}

We used a simple model to address the two issues that create the  
angular-momentum crisis of galaxy formation within the CDM scenario. 
One, the angular momentum catastrophe 
is that the spin of the baryonic 
component in galaxies is typically comparable
to that of the dark-matter halo (in bright galaxies) or even larger 
(in dwarf galaxies), while simple theoretical arguments and simulations
involving gas dynamics predict a significant spin loss by the baryons
due to over-cooling in merging halos.
The other, the mismatch of angular-momentum profiles,
is that the baryons in each galaxy tend to lack
the low-$j$ tail (and the high-$j$ tail) of the distribution as predicted 
by simulations for the dark halos.

In an earlier paper \citep{mds:02}, we showed that a simple algorithm, based
on adding up the orbital angular momenta of the mergers in
random realizations of merger histories, can successfully reproduce 
the distribution of spins among dark-matter halos as measured in 
N-body simulations of the $\lambda$CDM cosmology.
We showed here that an extension of such a model also reproduces the 
characteristic angular-momentum profile, i.e., the distribution of 
specific angular momentum within halos.
This provided the basic tool for addressing a possible resolution to the spin
problems by incorporating the effects of supernova feedback on the
gas in halos before they merge into bigger halos 
--- a process we have termed spin segregation. 

A simple analysis of how the orbital angular momentum in a merger turns into
a spin profile suggested how feedback effects in the satellite before the 
merger event can eliminate the problem of spin loss.  The 
effective size of the gas component within the incoming 
satellite determines its tidal stripping position in the 
halo and thus the final spin that it will be left with after the merger.
The finding, using the orbital-merger model, that the low-end tail
of the $j$ distribution originates in many minor mergers, that
tend to cancel each other's angular momentum, provided
a possible solution to the spin-profile mismatch problem.
The blowout of gas from small incoming halos
would eliminate the low-$j$ tail of the baryon distribution in the 
merger product, as observed.
The blowout has a particular strong effect in dwarf galaxies, 
because they are made of smaller progenitors that tend to loose 
more of their gas.  This results in 
a higher spin parameter for the baryons than the dark matter, 
as observed.

We constructed a simple semi-analytic model for simulating this process
using a simplified model for the effects of feedback as a function
of halo mass, including heating and blowout.
For a given choice of exponent $\gamma_2$, the model has one free parameter, 
the characteristic halo virial velocity $\vfb$ 
for which the energy inputted to the IGM by supernova
is sufficient to counter the effect of cooling.
By matching the low baryonic fraction in the dwarf galaxies observed by BBS,
we found that for $0.8 \leq \gamma_2 \leq 3$ the characteristic velocity
has to be in the range $60 \leq \vfb \leq 130 \kms$,
which falls within the range of theoretical predictions (e.g., DS). 
We then found considerable agreement between the model predictions and 
the observed data for both the distribution of the spin parameter
and the angular-momentum profile of baryons in dwarf and bright galaxies.
We also noted that the same basic model may explain the other unresolved
issues concerning angular momentum in galaxies, such as the spread in
observed disk sizes and the identification of halos that
form late type galaxies. The success of the model in matching several
independent observations indicates that this simple 
model indeed captures the main features necessary for a full scenario 
involving feedback and mergers that can resolve the spin crisis 
in more detail. The next natural step should be to incorporate
a more sophisticated feedback recipe into the model using the full 
machinery of semi-analytic models of galaxy formation.
This will be a step towards the long-term goal of implementing feedback in
full-scale hydrodynamical cosmological simulations.

Our work is an attempt to resolve the angular-momentum problems within the
standard framework of CDM cosmology, which is so successful on large scales,
using the effects of feedback which we know must occur.
Another approach is to appeal to a different cosmological scenario, WDM,
where the dark-matter particles are ``warm" rather than ``cold" 
\citep{hogan:99,hd:00,pp:82}. 
WDM is less robust than CDM because it requires fine-tuning of a new parameter,
the particle mass, which is constrained to be $\simeq 1~keV$.
Still, it is worth investigating since it may remedy the angular-momentum 
problems without appealing to strong feedback effects. 
The main distinguishing feature of the WDM 
scenario is that the formation of small halos is significantly suppressed, 
such that the validity of the explicit picture of halo buildup 
by the hierarchical build up of 
small halos becomes limited. Despite this difference from CDM,
an N-body simulation of WDM \citep{bkc:01} indicates
that the angular-momentum properties of halos remain basically unchanged.
This is not very surprising because in \citet{mds:02} we found that the same
angular-momentum properties can also be interpreted as a general result
of tidal-torque theory, independent of the explicit picture of mergers 
(see also BD).
In the absence of small halos at early times, one may expect less
over-cooling in halos before they merge into other halos and thus less
angular-momentum loss by the baryons. 
Indeed, hydrodynamical simulations of this scenario \citep{sld:01}
indicate that the angular momentum catastrophe is significantly remedied.
However, the angular-momentum profile mismatch is still expected to be valid 
in WDM, and, in the absence of small halos, the feedback effects are weaker
and may not be enough for resolving the problem.
Thus the $j$-profile mismatch may be a crucial discriminator between
solutions to the problems of CDM.

It is clear that some sort of heating 
may provide the cure for another problem of CDM, the dwarf satellite 
problem \citep{klyp:99b,moore:99a}, where the predicted number of dwarf 
halos is much larger than the observed number of dwarf galaxies.
\citet{bkw:00} demonstrated that cosmological photo-ionization 
due to feedback from UV sources such as early stars and quasars
can solve this problem. Feedback from supernova would qualitatively
have a similar effect.
While the number of dwarf satellites is automatically suppressed in WDM,
it seems that the inclusion of minimum realistic feedback effects 
would reduce the predicted number of galaxies to significantly below the 
observed number, and thus be an overkill (J. Bullock, private communication).

Furthermore, it is becoming clear (Dekel et al., in preparation)
that the key elements of our model, 
namely the tidal effects in mergers and the feedback effects in small
halos, are also very relevant in understanding and resolving the third
problem of CDM. This is the cusp/core problem,
where the halos in simulations typically show steep cusps in their inner 
profiles \citep{nfw:95,moore:99a}, while observations indicate flat cores at 
least in some low-surface-brightness galaxies \citep{bmr:01}.  
An analysis of tidal effects explains the necessary formation of an asymptotic
cusp in halos as long as satellites continue penetrating into the halo center. 
Feedback effects may puff up small satellites and prevent them
from penetrating the core, thus allowing a stable core.
These studies together indicate that our model indeed grasps the 
relevant elements of the complex processes involved, and that
feedback effects may indeed provide the cure to all three major problems 
of galaxy formation in CDM.


This research has been supported by the Israel Science Foundation
grant 546/98, by the US-Israel Binational Science Foundation
grant 98-00217, and by the German-Israeli Science Foundation
grant I-629-62.14/1999.
We thank James Bullock and Frank van den Bosch for stimulating discussions.
We also thank Frank van den Bosch for providing us with the BBS data in 
\Fig{amd}.

\bibliographystyle{mn2e}         
 
\bibliography{me,t,spin,spin2} 

\end{document}